\documentclass[aps,prb,preprint,showpacs]{revtex4-1}
\usepackage{graphicx}
\usepackage{amssymb,amsfonts,amsmath}
\usepackage{bbm}
\usepackage{dcolumn}
\usepackage{color}
\usepackage{subfig}
\usepackage{rotating}
\usepackage{verbatim}

\begin{document}

\title{Phase Diagram of Carbon Nickel Tungsten: Superatom Model}
\author{Sanxi Yao}
\author{Qin Gao}
\author{Michael Widom}
\affiliation{Department of Physics, Carnegie Mellon University, Pittsburgh PA 15213}
\author{Christopher Marvel}
\author{Martin Harmer}
\affiliation{Department of Materials Science and Engineering, Lehigh University, Bethlehem, PA 18015}

\date{\today}

\begin{abstract}
Carbon solubility in face-centered cubic Ni-W alloys and the phase diagram of C-Ni-W are investigated by means of first principles calculations and semi-grand canonical Monte Carlo simulations. With density functional theory (DFT) total energies as fitting data, we build a superatom model for efficient simulation. Multi-histogram analysis is utilized to predict free energies for different compositions and temperatures. By comparing free energies of competing phases, we are able to predict carbon solubility and phase diagrams of C-Ni-W at different temperatures. A simple ideal mixing approximation gives qualitatively similar predictions.
\end{abstract}


\maketitle

\section{Introduction}
Nickel-tungsten (Ni-W) alloys have gained interest recently due to their unusual mechanical, electrical and corrosion properties, including high hardness, melting temperature and corrosion resistance \cite{Reed2008,Eliaz2005,Yao1996}. W atoms dissolve in the fcc matrix of Ni~\cite{Raghavan2009}, even though the stable pure W structure has a bcc lattice~\cite{Eina1997}. Owing to their limited grain growth at elevated temperature, Ni-W alloys are considered to be thermally stable~\cite{Detor2007}. Three Ni-W phases are experimentally reported~\cite{Naidu1986}: Ni$_4$W, NiW and NiW$_2$. However, the NiW and NiW$_2$ phases are actually carbon nickel tungsten (C-Ni-W) phases CNi$_6$W$_6$ and CNi$_2$W$_4$~\cite{Cury2009}. A recent theorectical study~\cite{Schin2014} reports a tendency for tungsten atoms to order along chains in the $\langle$100$\rangle$ direction to form a Pt$_8$Ti phase, and also predicts possible stable structures with W compositions larger than 20at. $\%$. The Pt$_8$Ti-type structures is also suggested by x-ray scattering off the Fermi surface~\cite{Maisel2013}. Considering common impurities, carbon has strong interaction with Ni-W~\cite{Marvel2015}, leading to formation of C-Ni-W phases at low carbon concentration.

To study carbon solubility in FCC Ni-W, we calculate the energies of Ni-W structures and fit an energy model. We then perform semi-grand canonical Monte-Carlo simulation to study the phase stability of Ni-W as well as C-Ni-W with carbon at octehedral interstitial sites. We also make predictions about the tungsten solubility range as a function of temperature for the Ni-W binary. By studying the effect of impurities like carbon, we propose a explanation to why the Ni$_8$W phase, which is stable from calculation, is not observed in experiment, and then show our predicted phase diagrams of the C-Ni-W ternary at several temperatures, with both Monte Carlo simulation and ideal mixing approximation methods.

\section{Methods}
\subsection{First principle calculation and convex hull of Ni-W}
We use the density functional theory-based Vienna ab-initio simulation package (VASP)~\cite{Kresse93,Kresse94,Kresse961,Kresse962} to calculate total energies within the projector augmented wave (PAW)~\cite{Blochl94,Kresse99} method, utilizing the PBE generalized gradient approximation~\cite{Perdew96,Perdew97} as the exchange-correlation functional.  We construct a variety of supercells, about 440 Ni-W binary structures in all, which we relax with increasing $k$-point meshes until convergence is reached at the level of 0.1 meV/atom, holding the plane-wave energy cutoff fixed at 270 eV. For phonon calculation, we increase the cutoff energy to 330 eV and use a smaller energy tolerance to ensure better convergence. A few special C-Ni-W structures are made and calculated to evaluate the nearest neighbor carbon-carbon and carbon-tungsten repulsion. For the C-Ni-W ternary the cutoff energy is 400 eV.

\begin{figure}[h]
\centering
\includegraphics[width=3.3\linewidth,trim={0 0 0 0},clip]{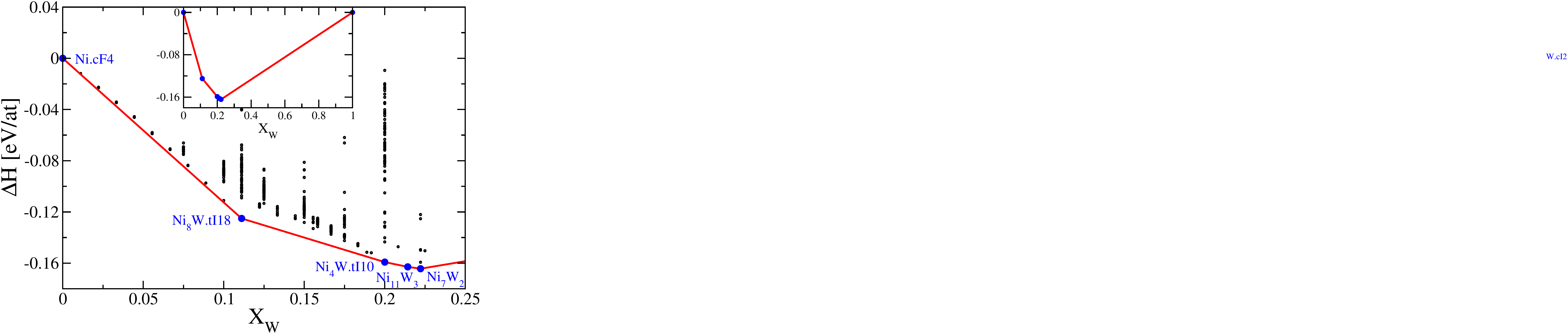}
\caption{Enthalpy of formation $\Delta H$ of FCC-based Ni-W structures (black and blue) calculated with VASP and the convex hull (red lines) in the range with tungsten composition less than 25$\%$. Solid blue circles shows the stable structures on the convex hull. Inset shows the full composition range.}
\label{fig:convexhull}
\end{figure}

Fig.~\ref{fig:convexhull} shows our first principles calculation result of the Ni-W binary, the entalphy of formation $\Delta H$ {\em vs.} the W composition, where $\Delta H$ is the energy of the structure with respect to pure elemental ground state structures, Ni.cF4 and W.cI2. The black and blue circles show the structures that we calculated. The red lines show the convex hull formed by stable structures. Any structure with $\Delta H$ above the convex hull will decompose into the end-point structures on the convex hull. We define $\Delta \Delta H$ as $\Delta H$ of the structure relative to the convex hull at the same compositon. For the stable binary structures, Ni$_8$W.tI18 and Ni$_4$W.tI10 have 11 at.$\%$ and 20 at.$\%$ W, respectively. Two other structures suggested by reports in the literature~\cite{Schin2014} are also on the convex hull, with W content slightly higher than 20 at.$\%$, consistent with their conclusion that Ni-W forms stable and metastable compounds over a wide composition range. These 4 stable structures all share a feature of columns formed by W atoms along $\langle 100\rangle$ directions. Variants made by displacing complete W columns also have small positive $\Delta \Delta H$.

\subsection{Superatom model}
\begin{figure}[ht]
\centering
\subfloat[]{\includegraphics[width=0.4\linewidth,trim={6.5cm 2.5cm 6.5cm 2.5cm},clip]{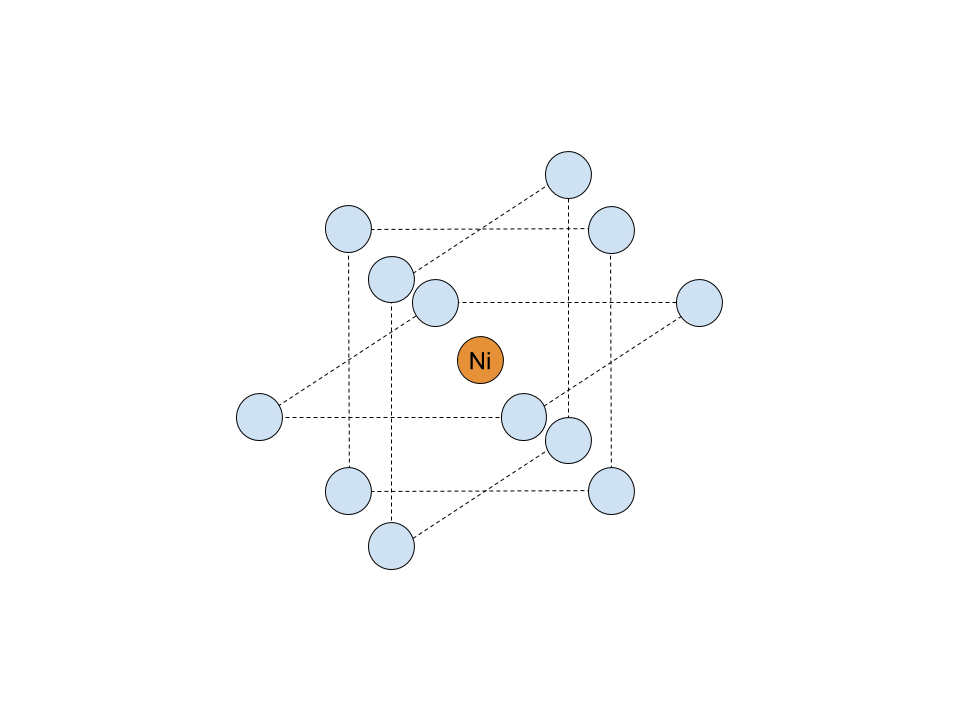}}
\subfloat[]{\includegraphics[width=0.4\linewidth,trim={6.5cm 2.5cm 6.5cm 2.5cm},clip]{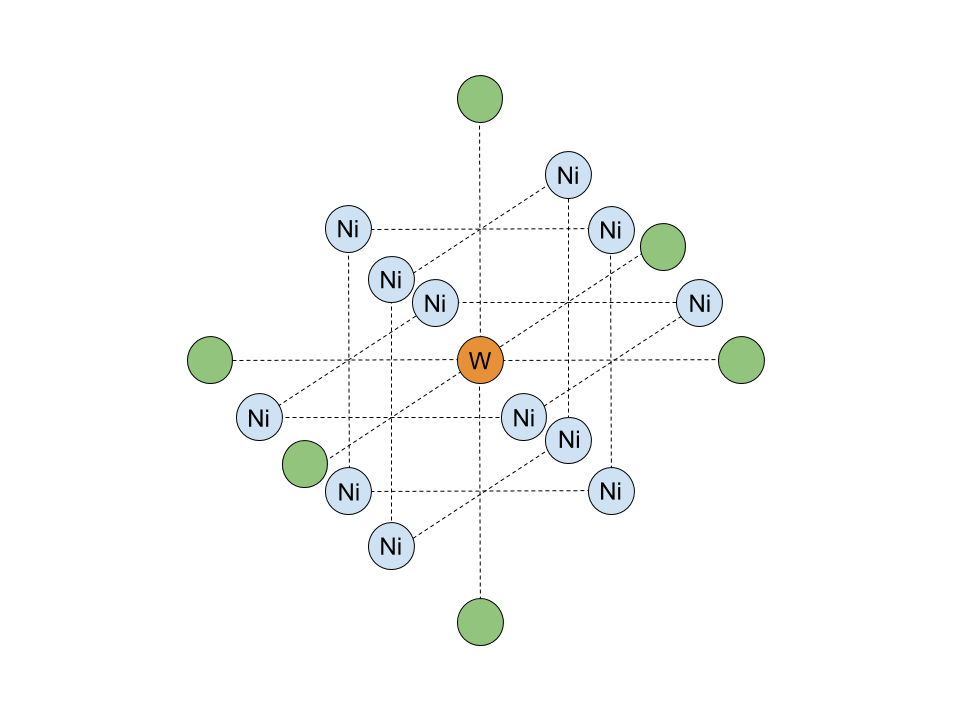}}
\caption{(a) superatom with central Ni atom and variable occupation of neighbors (blue). (b) superatom with central W atom, Ni first neighbors and variable occupation of second neighbors (green).}
\label{fig:superatom}
\end{figure}

We build a model to map FCC-based Ni-W structures to relaxed total energies. Our superatom model~\cite{Ektarawong14} includes local environments of Ni and W atoms, as well as longer-range pair interactions.  For the Ni-W binary on the FCC lattice, we first count the number of W atoms and the five shortest W-W pairs. Other pairs (Ni-Ni and Ni-W) are not considered since they are redundant with W-W pairs. Secondly, for each atom, we consider the environment around that atom up to the next nearest neighbor, and call it a superatom (Fig.~\ref{fig:superatom}). Ni-centered superatoms have variable occupation of 12 nearest neighbor sites resulting in 10 symmetry-independent cases. W-W near neighbor bonds are strongly unfavorable, so we define W-centered superatom to have only Ni as nearest neighbors. The 6 next-nearest neighbors have variable occupancy, resulting in 10 symmetry-independent types. Our superatom model includes complex many-body interactions that cannot be conveniently represented using a complete cluster expansion.

As a result, the superatom model contains twenty-six features: the number of W atoms, the numbers of five shortest W-W pairs, and the numbers from twenty types of super-atoms. The energy model is then expressed in the linear form:
\begin{align}
E(\vec{n})=\sum_{i=0}^{26}{\beta_in_i}=\vec{n}\cdot\vec{\beta}
\end{align}
Here $E$ is the energy per atom while $\vec{n}$ is the count of certain pairs or superatoms, divided by the total number of atoms. $n_0$ is always 1, corresponding to an overall energy shift, $n_1$ is W composition, $n_2$ to $n_6$ are the counts of W-W pairs scaled to per atom, and $n_7$ to $n_{26}$ are the counts of superatoms scaled to per atom.  Components of $\vec{\beta}$ are the corresponding fitting parameters.

\begin{figure}[ht]
\centering
\subfloat[]{\includegraphics[width=0.45\linewidth,trim={0 0 0 0},clip]{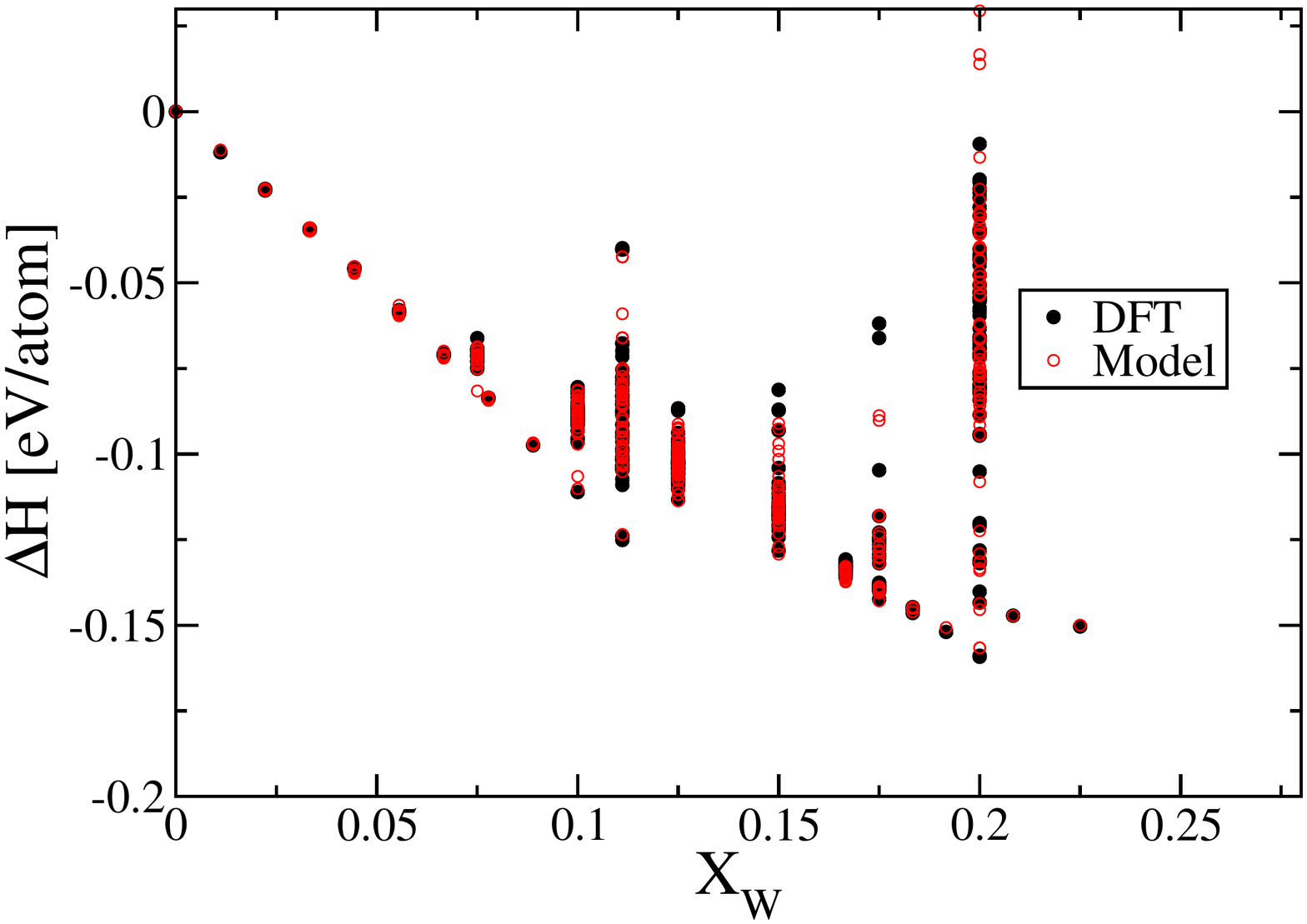}}
\subfloat[]{\includegraphics[width=0.45\linewidth,trim={0 0 0 0},clip]{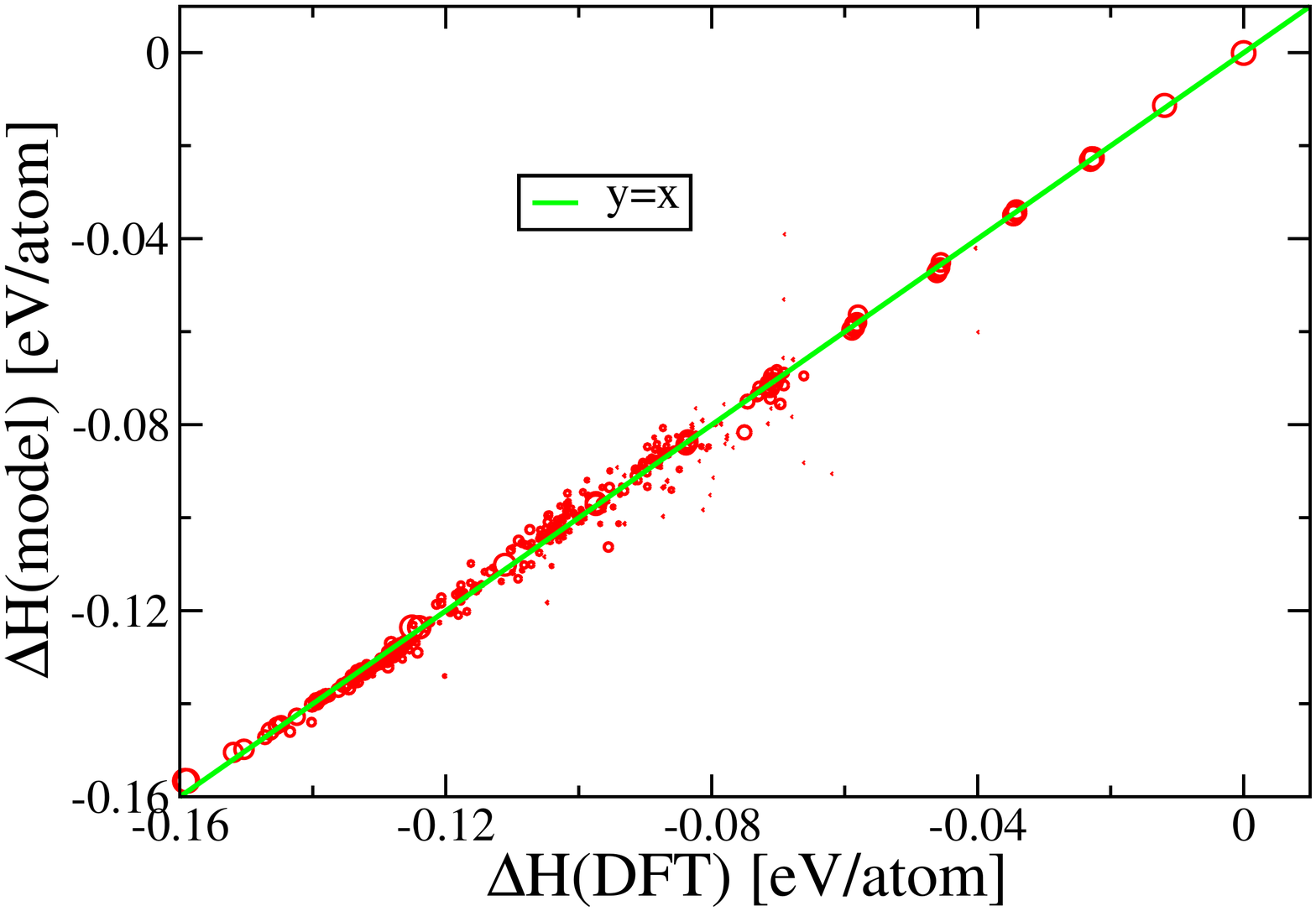}}
\caption{Comparison between DFT and superatom model. (a) enthalphy of formation $\Delta H$ as a function of W composition calculated via DFT (black circles) or predicted with the model (red circles). (b) Comparsion of DFT and superatom model, the size of circles indicates the weights in the fitting.}
\label{fig:fitting}
\end{figure}

Our fitting procedure minimizes the weighted-mean-square deviation of model energy from calculated DFT energy, where the weight is chosen as an exponential of $\Delta \Delta H$
\begin{align}
\beta^*=\arg\min_{\vec{\beta}}{\sum_\alpha{(e^{-{\Delta \Delta H_\alpha\over k_BT}}(\Delta H_\alpha - \vec{n}_\alpha\cdot \vec{\beta} ) )^2 }}.
\end{align}
Here $\Delta H_\alpha$ is the enthalpy of formation for structure $\alpha$, taking pure element structures as the reference points, and $\Delta\Delta H_\alpha$ is the energy above the convex hull. Our fit yields a 5-fold cross validation error around 2 meV/atom, corresponding to an uncertainty in temperature of about 23K.  The comparison between DFT and model is shown in Fig~\ref{fig:fitting}.

\subsection{Monte Carlo simulation and multi-histogram method}
We perform conventional Metropolis Monte Carlo simulations with 90 atoms in a 3$\sqrt 5\times 3\sqrt 5\times 1$ supercell, with 180 atoms (double along z axis), 270 atoms (triple along z axis) and 360 atoms (quadruple along z axis). This supercell was chosen to be commensurate with both Ni$_8$W.tI18 and Ni$_4$W.tI10 structures. The basic move is to randomly pick a site and change the species (Ni to W, or W to Ni). Moves are accepted or rejected according to the Boltzmann factor for the energy change $\Delta E$ and chemical potential change $\mu\Delta N_W$.  Here $\Delta N_W=-1$ when we pick a W atom and change it to Ni and $\Delta N_W=+1$ {\em vice versa}. The Boltzmann factor is $\exp({-(\Delta E-\Delta N_W\mu)/k_BT)}$ where $\mu\equiv \mu_W-\mu_{Ni}$.

For the Ni-W binary, where Ni and W sit at FCC lattice sites, we construct about 440 Ni-W structures with W composition ranging from 0 to 23 at.$\%$, and take the energies to fit an energy model for simulation. In the C-Ni-W ternary, C atoms sit at octahedral sites of the FCC lattice, surrounded by six nearest metal (Ni/W) sites. Carbon-carbon, and carbon-tungsten repulsion are added to the binary energy model. We then perform semi-grand canonical Monte-Carlo simulation to study the phase stability of C-Ni-W phase at different temperatures, holding fixed the total number of metal atoms and, in the ternary, the total number of carbon atoms.

Multidimensional replica exchange can help avoid getting stuck in some competing states. For a selection of simulation trajectories, suppose the $i^{th}$ one is at temperature $T_i$ and chemical potential $\mu_i$, with total energy $E_i$ and tungsten number $N_i$, and the $j^{th}$ one has the similar corresponding features with subscript $j$. The probability of occurence of these two trajectories is proportional to the corresponding Boltzmann factor
\begin{equation}
P_1=P_i(T_i,\mu_i)P_j(T_j,\mu_j)\propto e^{-(E_i-N_i\mu_i)/k_BT_i}e^{-(E_j-N_j\mu_j)/k_BT_j}.
\end{equation}
If we swap these two trajectories, so that trajectory $i$ takes $T_j$ and $\mu_j$, and trajectory $j$ takes $T_i$ and $\mu_i$, then the probability is
\begin{equation}
P_2=P_i(T_j,\mu_j)P_j(T_i,\mu_i)\propto e^{-(E_i-N_i\mu_j)/k_BT_j}e^{-(E_j-N_j\mu_i)/k_BT_i}.
\end{equation}
So the acceptance probability to interchange the trajectories is
\begin{equation}
P=P_2/P_1=e^{\Delta\beta \Delta E-\Delta N\Delta(\beta \mu)}
\end{equation}
where we define $\Delta \beta =1/(k_BT_i)-1/(k_BT_j)$, $\Delta E=E_i-E_j$, $\Delta N=N_i-N_j$ and $\Delta(\beta \mu)=\mu_i/(k_BT_i)-\mu_j/(k_BT_j)$.

We analyze the Monte Carlo results using a multidimensional variant of the multi-histogram method~\cite{Ferrenberg1988,Ferrenberg1989},	similar to the one-dimensional analysis in our previous work on the boron carbide system~\cite{Yao2015}.  At temperature T and chemical potential $\mu$, the histogram of configurations with energy $E$ and tungsten composition $x$, $H_{T,\mu}(E,x)$, can be converted into a density of states $W(E,x)=H_{T,\mu}(E,x)$exp$((E-\mu xN)/k_BT)$ where $N$ is the total number of atoms. Here we fix $T$, and combine the histograms from different chemical potentials $\mu_i$ to evaluate the density of states. Then we can evaluate the partition function
\begin{equation}
Z(T,\mu)=\sum_{E,x} W(E,x) e^{-(E-\mu xN)/k_BT},
\end{equation}
which is accurate over a range chemical potentials covering all $\mu_i$'s, as long as the histograms of nearby $\mu$'s overlap. Composition $x_W$ can be obtained by differentiating the free energy $F=-k_BT\ln{Z}$ with respect to $\mu$.

\section{Results and discussion}

\subsection{Ni-W binary simulation}
We perform Ni-W binary simulations for systems with 90, 180, 270 and 360 atoms, at different temperatures and chemical potentials, supplemented with two-dimensional replica exchange in T and $\mu$ to reach better equilibrium. By comparing the stable structures on the convex hull, Ni.cF4, Ni$_8$W.tI18, Ni$_4$W.tI10 and W.cI2, including temperature dependent corrections from vibrational free energy and electronic free energy, we determine that the chemical potential $\mu$ should be in a range of $-1.1$eV through $+0.16$eV at T=500K and change to $-1.0$eV through $+0.11$eV at T=1200K. We then choose a selection of $\mu$'s that covers that range.

\begin{figure}[ht]
\centering
\subfloat[]{\includegraphics[width=0.45\linewidth,trim={0 0 0 0},clip]{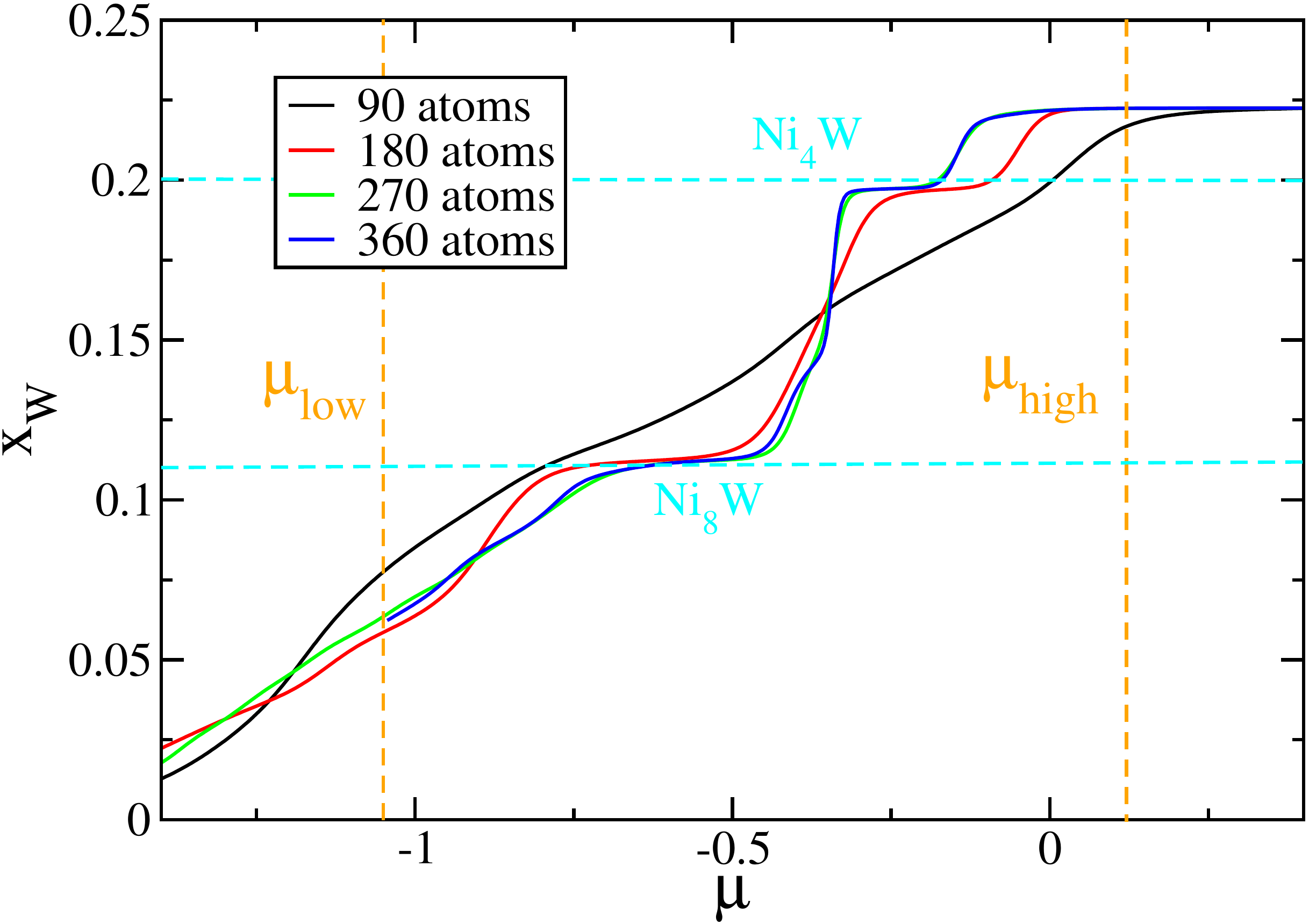}}
\subfloat[]{\includegraphics[width=0.45\linewidth,trim={0 0 0 0},clip]{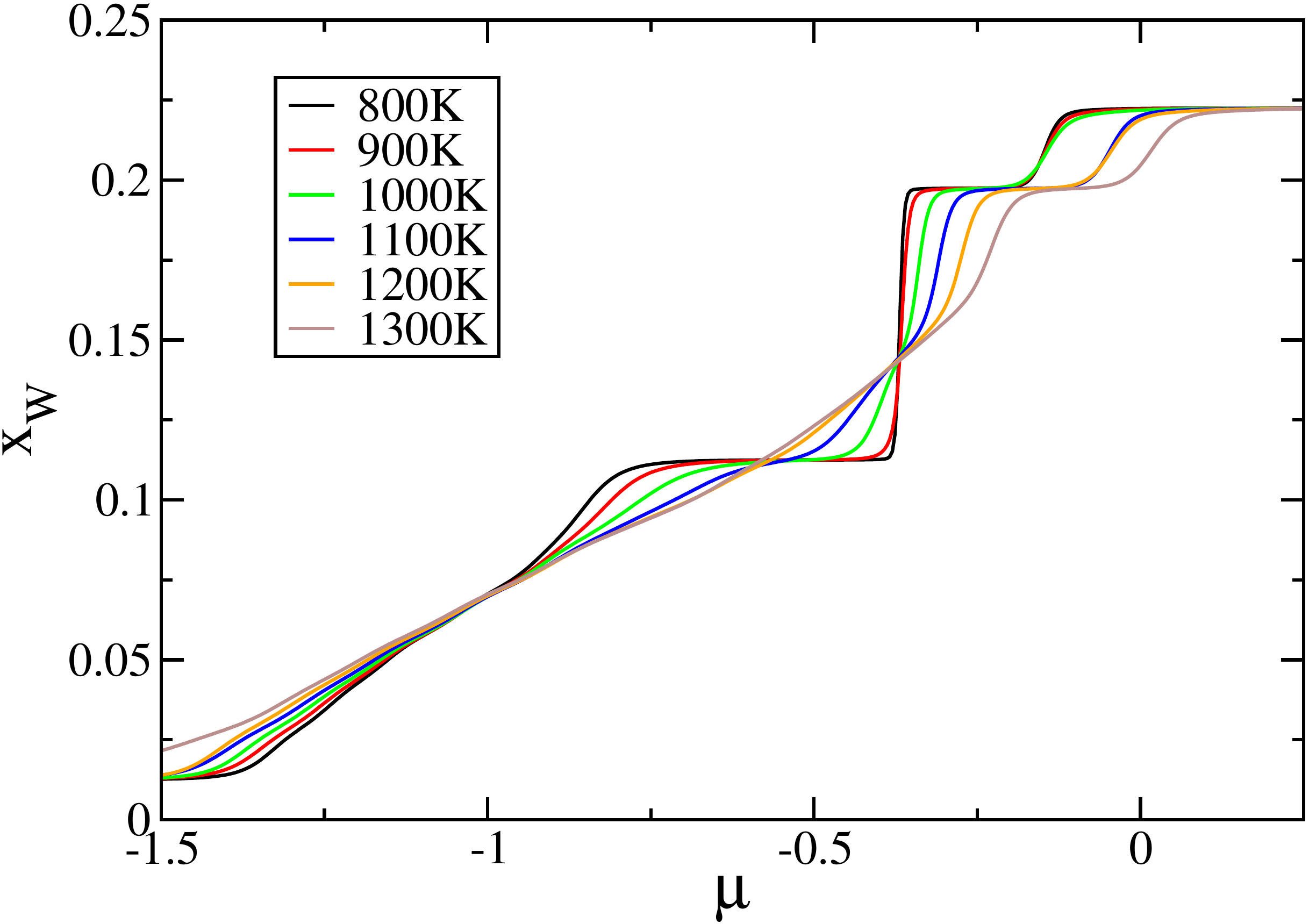}}
\caption{(a) Tungsten composition $x_{\rm{W}}$ as a function of $\mu$ for different system sizes at T=1000K, vertical orange dash line shows the real chemical potential limit at this temperature. (b) For system with 270 atoms, $x_{\rm{W}}$ as a function of $\mu$ at different temperatures.}
\label{fig:muxw}
\end{figure}

Fig.~\ref{fig:muxw} shows the predicted tungsten composition $x_{\rm{W}}$ as a function of $\mu$ for different system sizes (a) and at different temperatures (b). At T=1000K, the system with 90 atoms may be a little too small to show distinguishable Ni$_8$W and Ni$_4$W phase, while in larger systems those phases cover a range of $\mu$. The $x_{\rm{W}}$ curves for system with 270 and 360 atoms are overlapping, indicating that the finite size effect diminishes.

Fig,~\ref{fig:muxw}(b) shows $x_{\rm{W}}$ for a system of 270 atoms ($3\sqrt 5\times 3\sqrt 5\times 3$ supercell). As temperature increases, the Ni$_8$W phase begins to disorder around 1100K, while Ni$_4$W remains ordered until 1300K, but shifts towards high $\mu$. The high $\mu$ end, which is between 0.11eV and 0.16eV and temperature dependent, always goes beyond 20 at$\%$ of $x_{\rm{W}}$. Our first principles calculation in Fig.~\ref{fig:convexhull} displays some stable structures beyond Ni$_4$W, supporting this result.

\begin{figure}[h]
\centering
\includegraphics[width=0.9\linewidth,trim={0 0.8cm 0 2cm}]{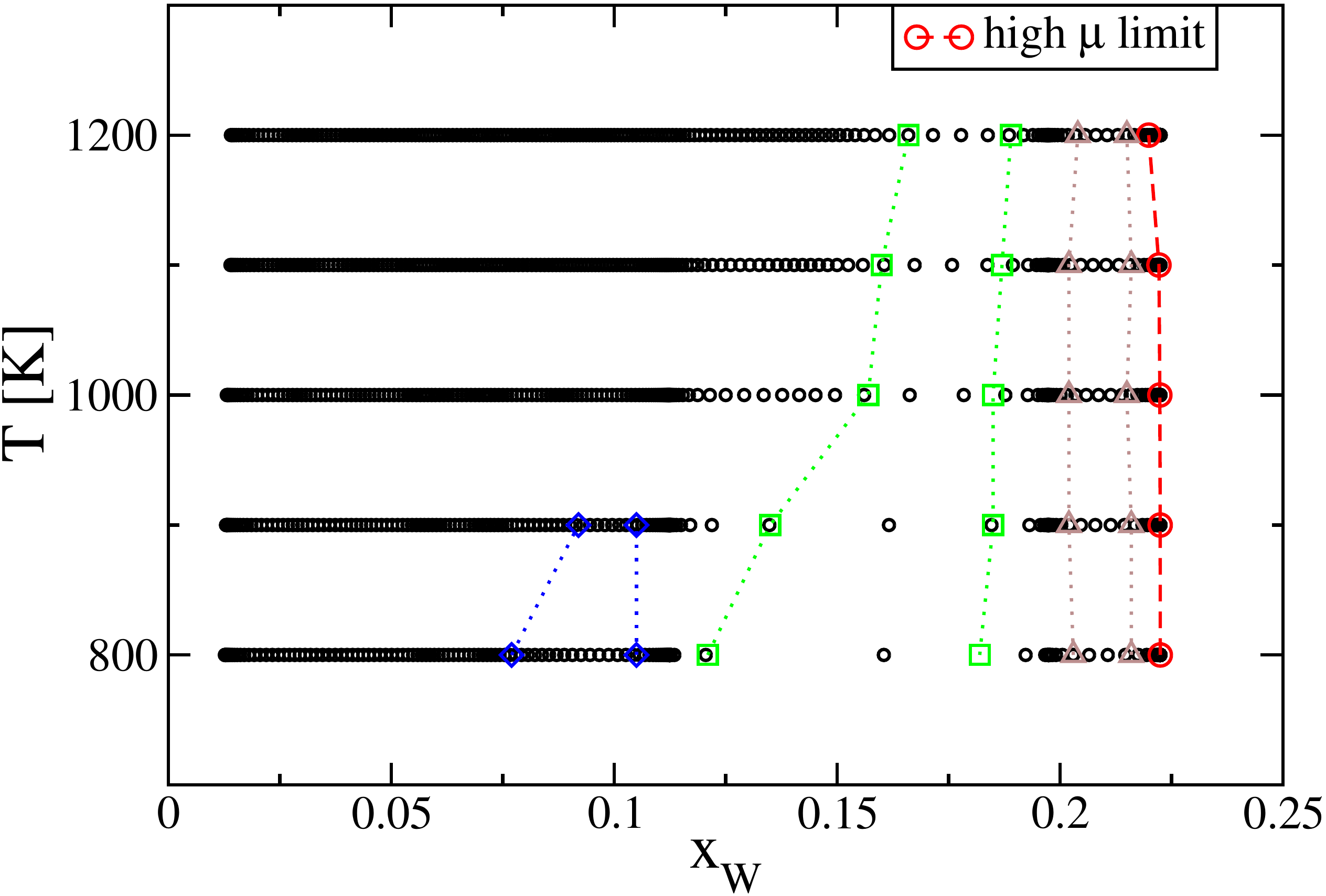}
\caption{Ni-W phase diagram in the $x_{\rm{W}}$-T plane for systems containing 270 atoms. Circles are calculated values on a uniformly spaced grid of chemical potential $\mu$. Red dashed line shows the real high $\mu$ limit. Blue dotted lines and green dotted lines shows the locus of peaks of $d^2x_W/d\mu^2$ in two regions. The peaks indicate the possible boundaries of two coexisting phases.}
\label{fig:niwphase}
\end{figure}

Figure~\ref{fig:niwphase} shows the simulated binary phase diagram for systems with 270 atoms. Each point represents our predicted $x_{\rm{W}}$ at one particular temperature and chemical potential. A micibility gap between Ni$_8$W and Ni$_4$W is shown by the dotted green lines.  The Ni$_8$W phase starts to merge with Ni$_4$W at temperature above 1000K. There is also a small gap between Ni$_4$W and phase beyond Ni$_4$W, shown by the grey dotted lines. Note that the calculation extends beyond the real high $\mu$ limit, which is shown as the red dashed line.

\subsection{Ni-W simulation with carbon}
Experimental studied conducted at Lehigh University have covered many aspects of nanocrystalline Ni-W alloys~\cite{Christ2016}, including grain growth, thermal stability, and the effect of processing. Their study quantified the impurity contentration of carbon, oxygen and sulfur in electron deposited alloys with hot extraction analysis, and determined that both carbon and oxygen concentrations reach up to 0.5 at.$\%$, while sulfur can be ignored relative to carbon and oxygen. Other light elements like nitrogen can also be present. Treating other impurities similarly as carbon, in affecting the thermal stability of C-Ni-W, we simulate up to 3.5 at.$\%$ total impurity concentration.

We add carbon at octahedral interstitial sites in the FCC lattice then conduct semi-grand canonical simulations where the number of carbon atoms is fixed, while the numbers of nickel and tungsten atoms are varied with their total number constant. There are two additional parameters in the energy model involving carbon: the energy of the shortest C-W bond and the energy of the shortest C-C bond. Both are repulsive and estimated from DFT calculation with the values 0.583 eV/bond and 0.302 eV/bond with respect to C-Ni bonds, respectively.

\begin{figure}[ht]
\centering
\subfloat[]{
\includegraphics[width=0.45\linewidth,trim={4cm 3.7cm 4cm 4.3cm},clip]{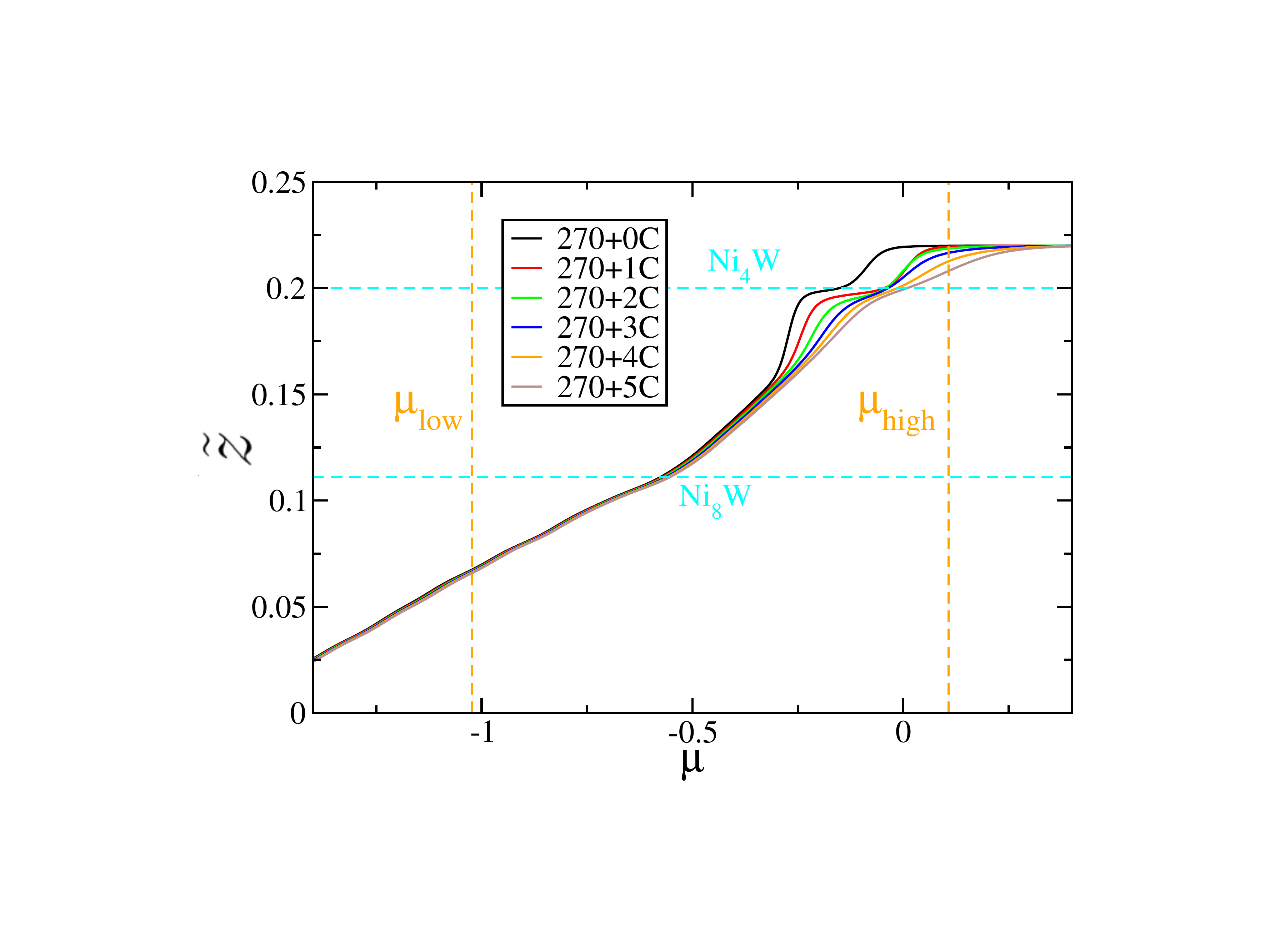}
}
\subfloat[]{
\includegraphics[width=0.45\linewidth,trim={4cm 3.7cm 4cm 4.3cm},clip]{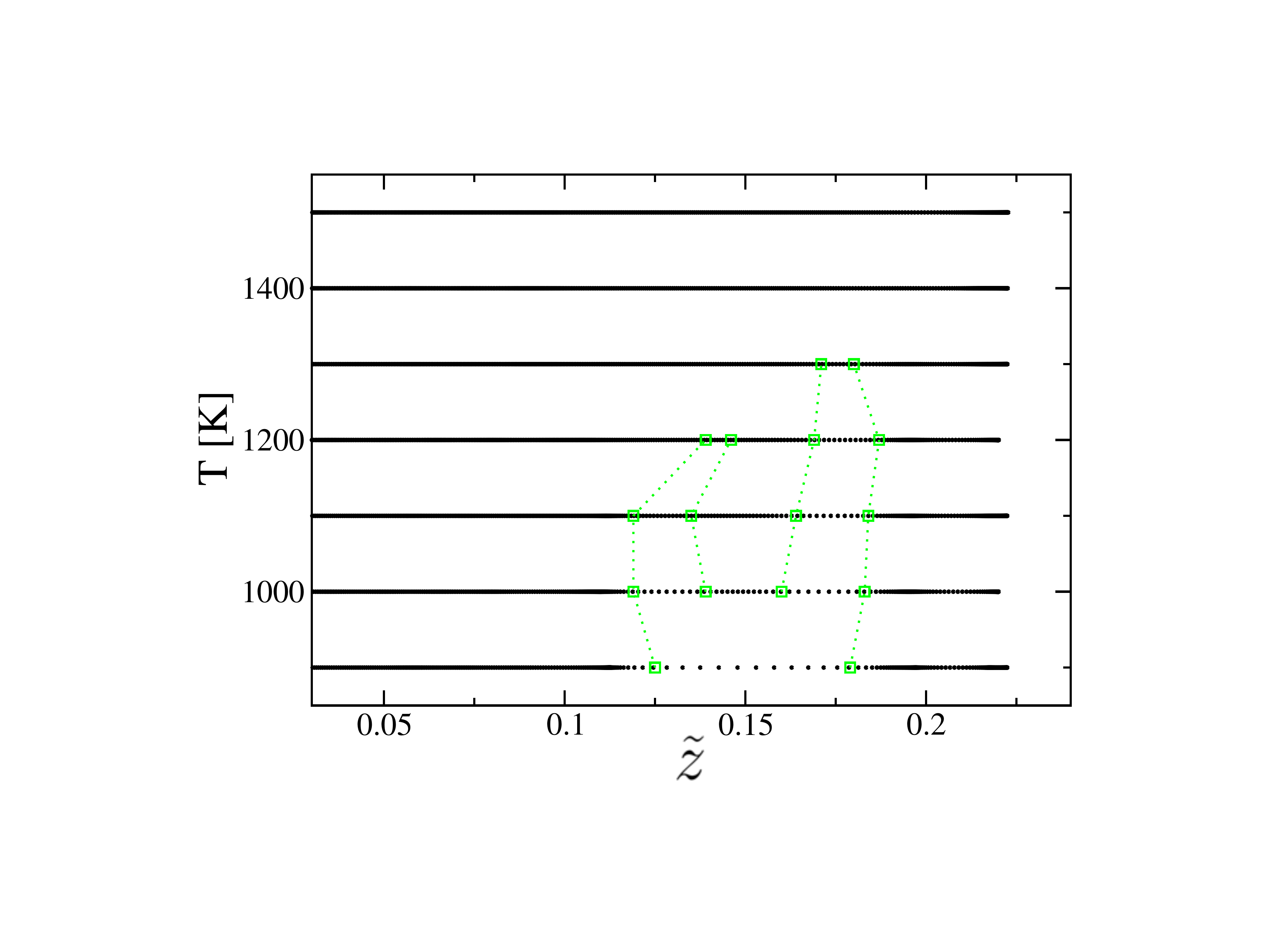}
}
\caption{(a) $\tilde{z}$ as a function of $\mu$ for the system with 270 Ni/W atoms with different number of carbons at T=1200K. (b) Ni-W phase diagram in the $\tilde{z}$-T plane for the system with 270 Ni/W atoms plus 2 carbon atoms. Green dotted lines show the locus of peaks of $d^2\tilde{z}/d\mu^2$. }
\label{fig:270C2}
\end{figure}

Figure~\ref{fig:270C2}a shows the predicted tungsten composition for the system with 270 metal atoms with different numbers of additional carbon atoms. Here we take the tungsten composition as the number of W atoms scaled by the total number of W and Ni atoms, without including the additional carbon atoms. We define this quantity as $\tilde{z}=z/(y+z)$ for the ternary phase C$_x$Ni$_y$W$_z$. At the high $\mu$ end, adding carbon decreases $x_{\rm{W}}$, dropping the high $\mu$ limit of the Ni-W phase to 20 at.$\%$. Fig.~\ref{fig:270C2}b displays the predicted phase diagram in the case with 270 Ni/W atoms and 2 carbon atoms, showing a gap between Ni$_8$W and Ni$_4$W (i.e. a solubility limit for the Ni-W solid solution of 11at.$\%$ W, consistent with experiment~\cite{naidu1986nickel}) at T below 1000K, and these two phases merge at higher temperature. It also shows that the Ni$_8$W phase disappears in presence of carbon, which could be a reason why the Ni$_8$W phase is not observed in experiments.

\subsection{C-Ni-W ternary phase diagram}
Based on the free energies obtained from the simulation results, and by comparing the free energy of C-Ni-W phase with that of competing phases, we can predict C-Ni-W ternary phase diagram. Besides elemental FCC Ni, BCC W and hexagonal C, we include  two special ternary structures CNi$_2$W$_4$.cF112 and CNi$_6$W$_6$.cF104, corresponding to the two previously proposed stable binaries NiW$_2$ and NiW~\cite{Naidu1986}, respectively. These two ternaries are predicted to be stable at finite temperature. Both CNi$_2$W$_4$.cF112 and CNi$_6$W$_6$.cF104, are derived from binary NiW.cF96, with carbon atoms sitting at octahedral interstitial sites, fully surrounded by W atoms (contrary to FCC solid solution, where C likes to be associated with Ni atoms). There are three types of Wyckoff positions in NiW.cF96 structures, with Ni sitting at two types of icosahedral sites while W is at Frank-Kasper type Z14 sites. W occupies one of the icosahedral sites in the case of NiW$_2$. Ni-W substitution is less costly in CNi$_2$W$_4$ than in CNi$_6$W$_6$, leading to a small range of solid solution around CNi$_2$W$_4$ at elevated temperature~\cite{fiedler1975ternary}. A different type of structure with stochiometry C$_7$Ni$_6$W$_{20}$.hP34 is calculated to be metastable, and can be stablized by phonon vibrational entropy at finite temperature. For the C-W binary line, our calculation shows the $\Delta H$ of CW$_2$ is high, about 68 meV/atom above the convex hull. Experimentally the CW$_2$ phase, which has intrinsic disorder, only appears above 1250$^{\circ}$C~\cite{okamoto08}.

\begin{figure}[ht]
\centering
\subfloat[]{\includegraphics[width=0.43\linewidth,trim={0 0 0cm 0},clip]{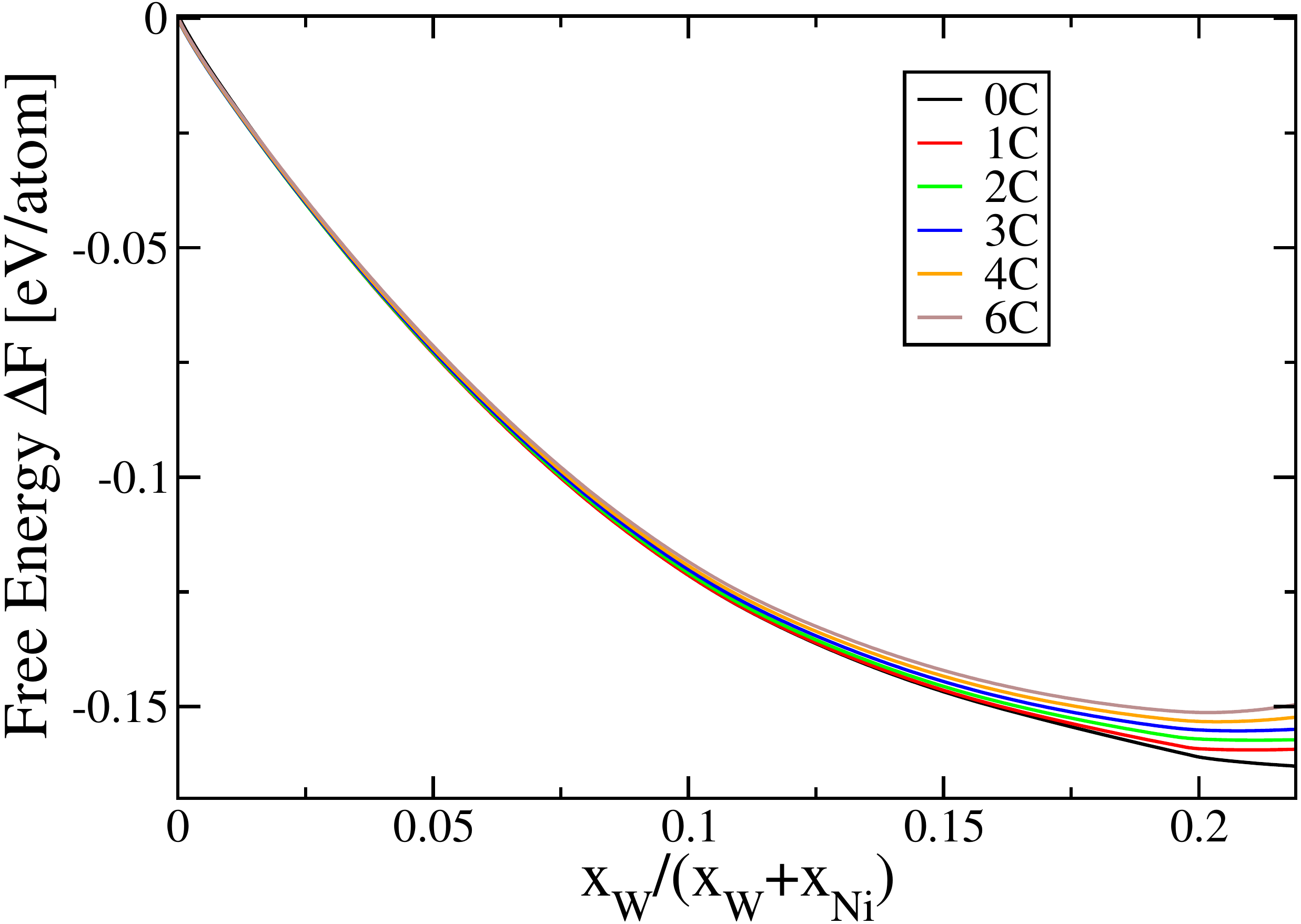}}
\hspace{1cm}
\subfloat[]{\includegraphics[width=0.3\linewidth,trim={0 0 0cm 0},clip]{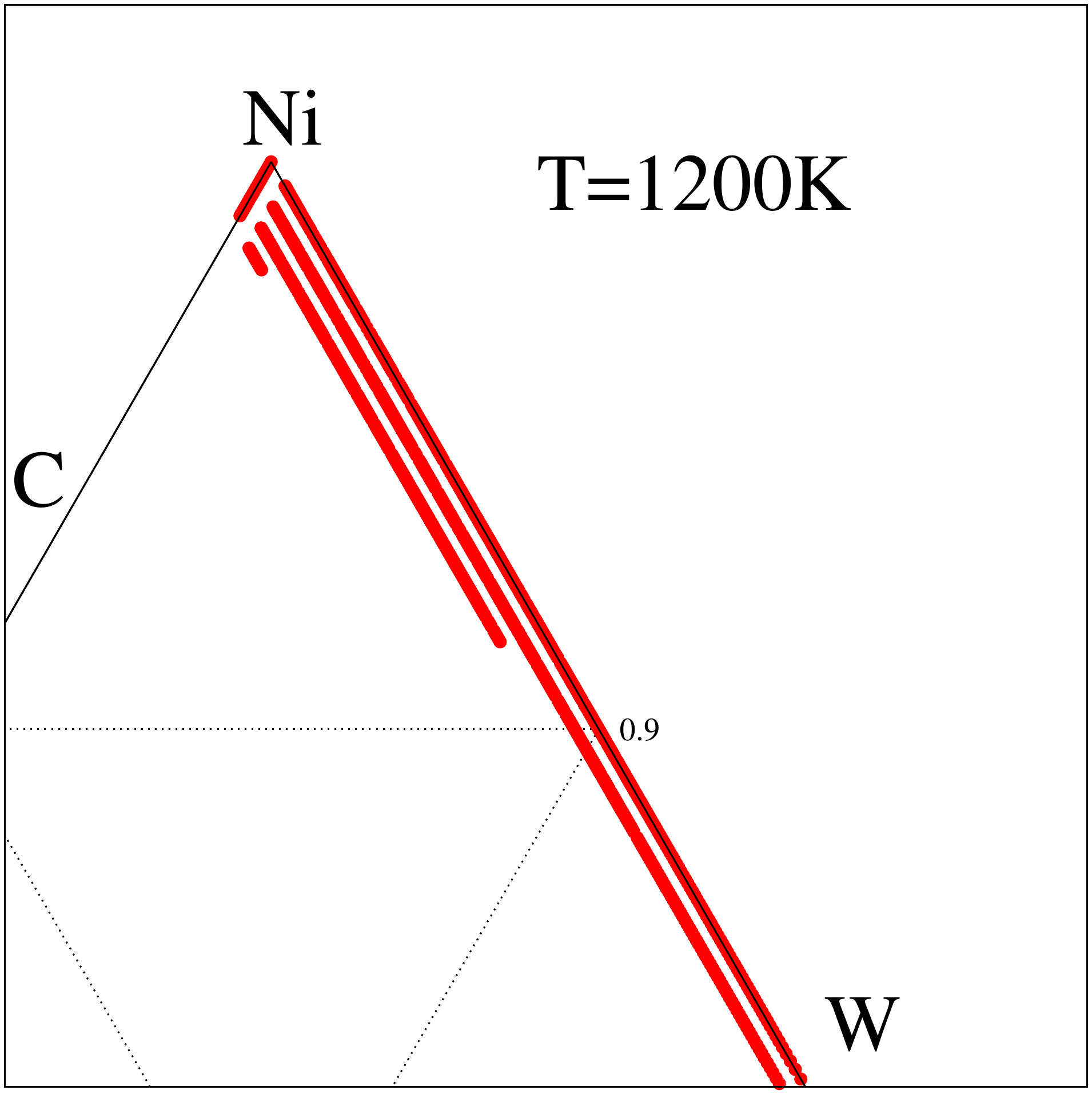}}
\caption{(a) Free energy predicted from simulation from system containing 270 Ni/W atoms, with various numbers of C atoms. (b) Enlargement of the predicted phase diagram at T=1200K at the high Ni corner. The red dots (connecting as lines) show the stable compositions lying on the convex hull, and occupy the region of stability of a single phase C-Ni-W FCC solid solution.}
\label{fig:how}
\end{figure}

Multiple histogram analysis of simulations predict free energies for systems with a fixed of carbon atoms. However, this free energy has an unknown reference as a function of the number of carbons.  For a specific number of carbon atoms, we predict the free energy $\Delta F$ as a function of $\tilde{z}=x_W/(x_W+x_{Ni})$. The free energy along the C-Ni line can be estimated with the linearly interpolated energy and ideal mixing approximation of entropy. Matching the lines of $\Delta F$'s with C-Ni line, we can put our predicted free energy on a unified relative scale, and determine the stable structures as those on the on the convex hull of $F$. Fig.~\ref{fig:how} shows an example from the system with 270 Ni/W sites. Part (a) shows free energies from simulations with different numbers of carbon atoms, while (b) shows the stable compositions of the C-Ni-W FCC solid solution as a set of red lines.

\begin{figure}[ht]
\centering
\subfloat[]{\includegraphics[width=0.45\linewidth,trim={0 0 9cm 0},clip]{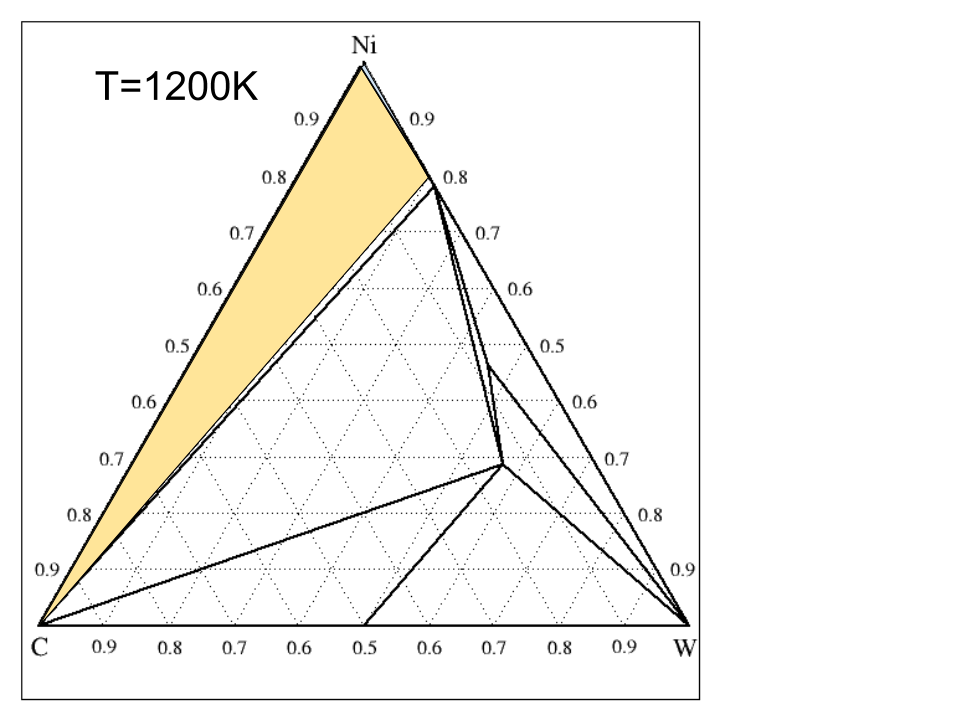}}
\subfloat[]{\includegraphics[width=0.45\linewidth,trim={0 0 9cm 0},clip]{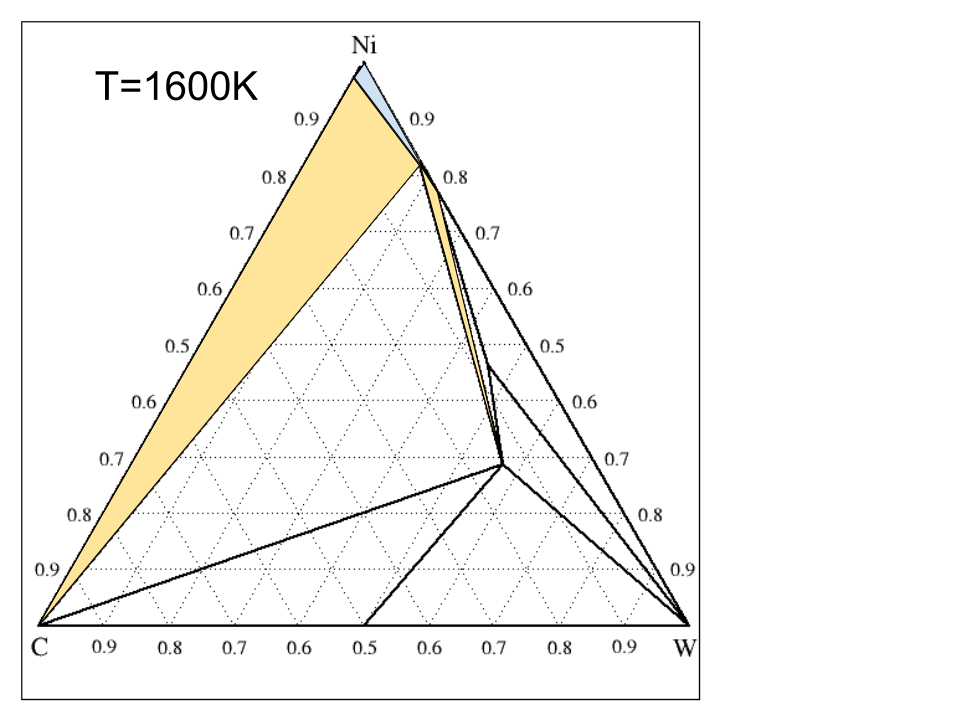}}
\caption{C-Ni-W ternary phase diagram at T=1200K (a) and T=1600K (b). The blue shows C-Ni-W FCC solid solution single phase region, the yellow is two-phase region and the white is three-phase region. Black lines show the convex hull.}
\label{fig:PDs}
\end{figure}

The predicted ternary phase diagrams at T=1200K and T=1600K are displayed in Fig.~\ref{fig:PDs}. At 1200K, there is a small region of FCC. As temperature increases to 1600K, the carbon solubility increases and the FCC phase region expands. The solubility expansion can also be seen along Ni-W binary line by comparing these two phase diagrams.

\subsection{Constrained ideal mixing approximation model}
A simple free energy model was also constructed, in which the energy is a piecewise linear function and the entropy is evaluated with an ideal mixing approximation under some constraints. For the fcc solid solution with interstitial carbon of composition C$_x$Ni$_y$W$_z$ $(x+y+z=1)$, the piecewise linear energy can be expressed as
\begin{equation}
E(x,y,z)=xE_{\rm{C}}+(y+z)E_{\rm{Ni-W}} \nonumber
\end{equation}
$E_{\rm{C}}$ is determined from C solubility in Ni~\cite{Singleton1990}. $E_{\rm{Ni-W}}$ is piecewise linear function of the convex hull of the phases Ni.cF$_4$, Ni$_8$W.tI18, Ni$_4$W.tI10 and W.cI2.

When estimating the entropy, we take a constrained ideal mixing approximation. Write C$_x$Ni$_y$W$_z$ as ${\rm{C}}_x{\rm{Ni}}_y{\rm{W}}_z=x{\rm{C}}+(y+z){\rm{Ni}}_{\tilde{y}}{\rm{W}}_{\tilde{z}}$ where $\tilde{y}=y/(y+z)$, $\tilde{z}=z/(y+z)$. Note that the number of octahedral interstitial sites equals the number of metal sites. Under this constraint, ({\em i.e.} the assumptions that W atoms have no C as nearest neighbors and that W atoms are widely separated) a fraction $6z$ of intersitital sites must be vacant. These assumptions are reasonable since the C-W bond has an enery of 0.583~eV/bond with respect to the C-Ni bond, and our fitting model yields positive values for the nearest and next nearest W-W pairs. Thus only $(y+z)-6z=(1-x)-6z$ interstitial sites are potentially occupied by C. Defining $\tilde{\mathbbm{1}}\equiv (1-x)-6z\ge 0$ as the available carbon site fraction, then we have $z\le \frac{1}{6}$, which implies we must keep $z\le \frac{1}{6}$ in this model. We have entropy
\begin{align}
-S/k_B=x\ln{x}+(\tilde{\mathbbm{1}}-x)\ln{(\tilde{\mathbbm{1}}-x)}+F_C(\tilde{z})(y\ln{\tilde{y}}+z\ln{\tilde{z}}).
\end{align}

Assuming that the entropy for Ni$_4$W vanishes, we include a factor $F_C(\tilde{z})=1-5\tilde{z}$ multiplying the Ni-W entropy since for Ni$_4$W $\tilde{z}=0.2$, and for pure Ni where $\tilde{z}=0$ we want to keep full entropy.

\begin{figure}[]
\centering
\includegraphics[width=0.4\linewidth,trim={5cm 0 4cm 0cm},clip]{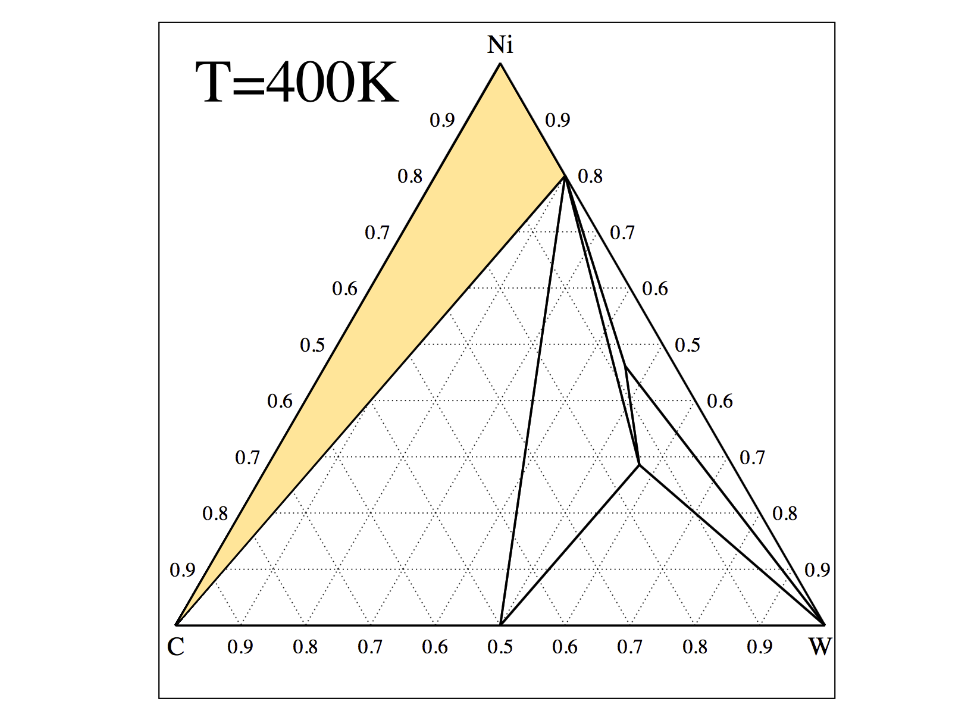}
\hspace{0.05cm}
\includegraphics[width=0.4\linewidth,trim={5.5cm 0 3.5cm 0cm},clip]{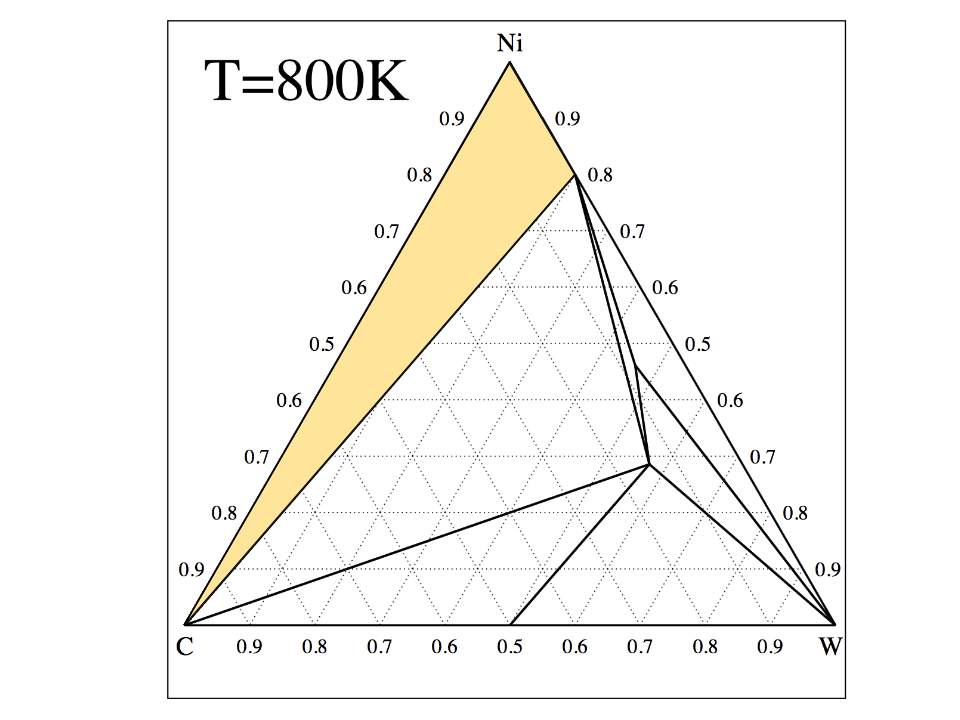}
\includegraphics[width=0.4\linewidth,trim={4cm 0 5cm 0cm},clip]{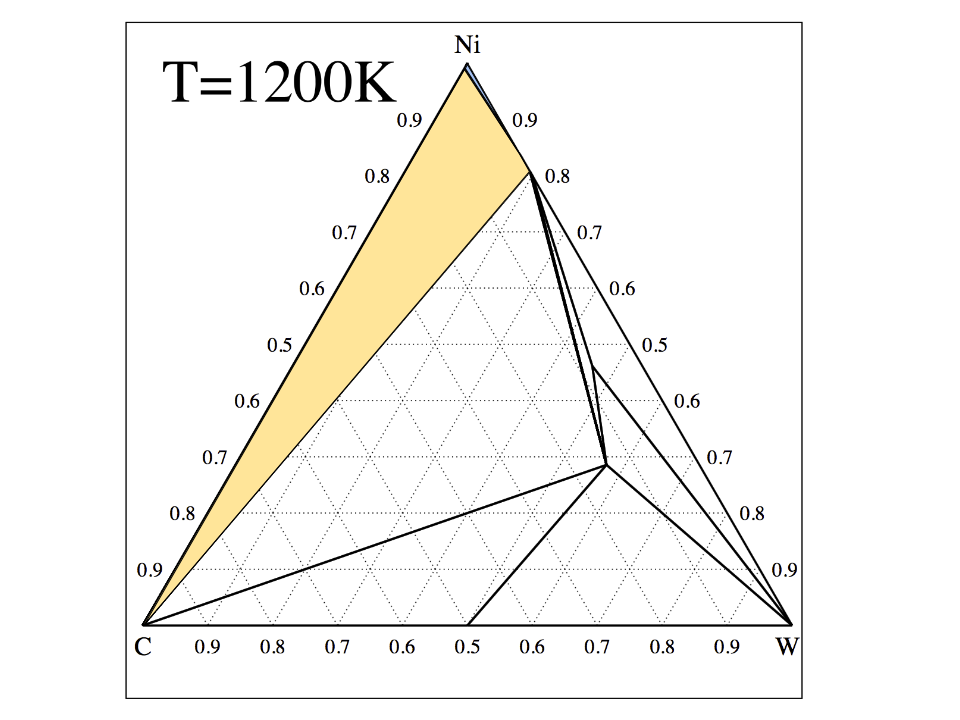}
\hspace{0.03cm}
\includegraphics[width=0.39\linewidth,trim={5.6cm 0.2cm 4.1cm 0cm},clip]{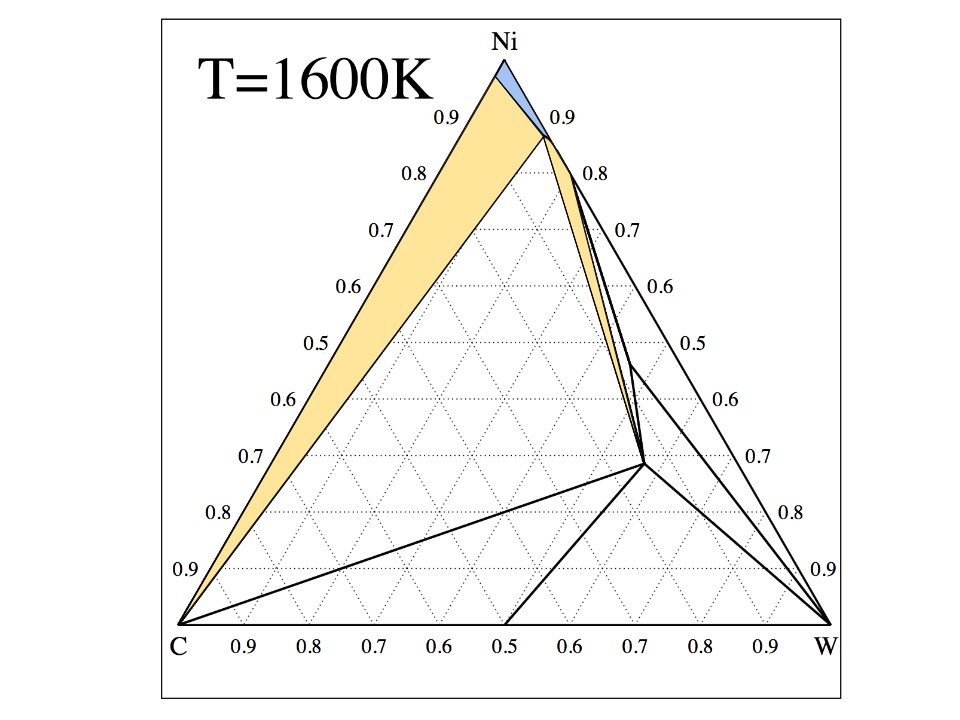}
\includegraphics[width=0.4\linewidth,trim={0.cm 0 8.8cm 0cm},clip]{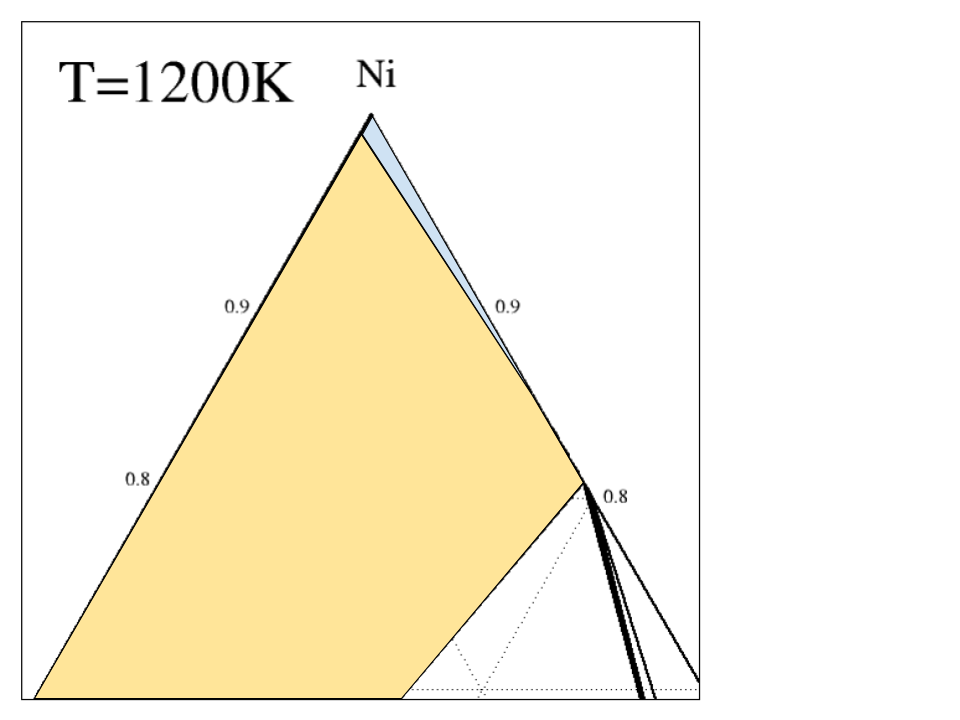}
\hspace{0.0cm}
\includegraphics[width=0.4\linewidth,trim={0.cm 0 8.8cm 0cm},clip]{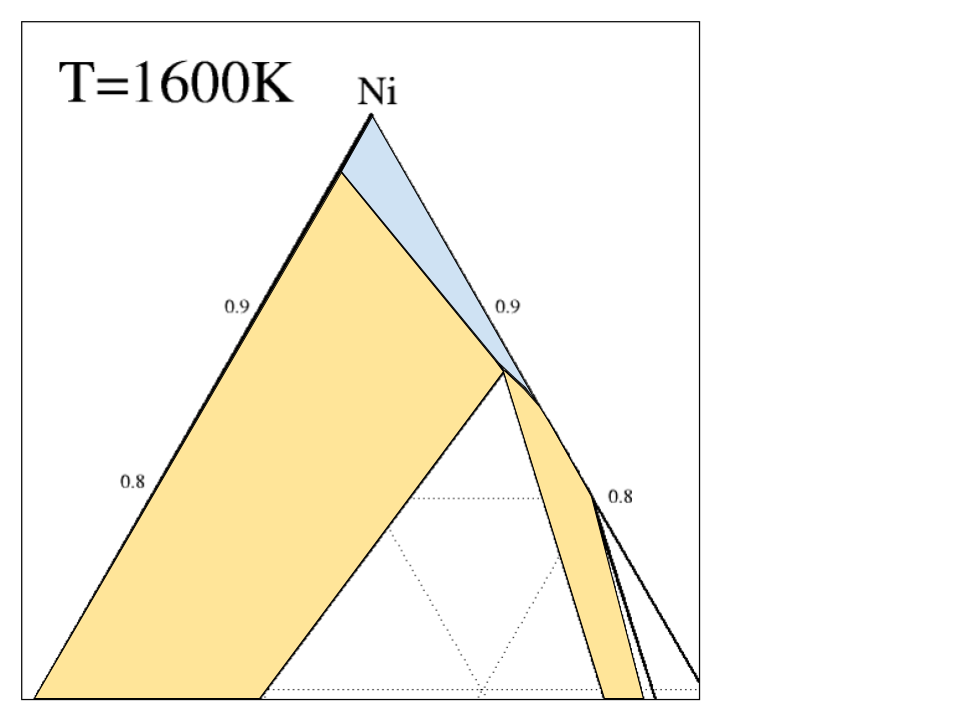}
\caption{C-Ni-W phase diagram at T=400K, 800K, 1200K and 1600K under the constrained ideal mixing approximation model. The two bottom figures show the high Ni corner at 1200K and 1600K, respectively.}
\label{fig:pdSimple}
\end{figure}

With this simplified model for the FCC solid solution, together with the free energies of other phases, we predict the phase diagram at various temperatures as shown in Fig.~\ref{fig:pdSimple}. This model also allows us to make predictions at lower temperature, which cannot be reached by simulations due to the difficulty of reaching true thermodynamic equilibrium at lower temperatures.

\section{Conclusion}
A superatom model was constructed for FCC solid solutions of Ni-W. This superatom model fits well to density functional theory total energies, serving as energy model for Monte Carlo simulation. With this model, we first studied the Ni-W binary phase diagram from 0-23 at.$\%$ tungsten composition. To study the ternary C-Ni-W solid solution, where carbon atoms are at octahedral interstitial sites, we augmented the model by including the carbon-tungsten, and carbon-carbon bond, nearest neighbor repulsion.

Monte Carlo simulations on a grid of temperatures and chemical potentials, together with replica exchange to attain better-equilibrium sampling, reveal a miscibility gap between Ni$_8$W and Ni$_4$W at low temperature; but a full solubility range at higher temperatures, above 1000K. At the high $\mu$ limit end the tungsten composition is slightly higher than 20 at$\%$. Including carbon atoms decreases the tungsten composition, especially at high $\mu$ end.  Understanding the experimentally reported~\cite{naidu1986nickel} binary solubility limit of 16\% W, however this falls naturally out of our constraint $z\le 1/6$ in the presence of impurities.

Comparing the free energies predicted from semi-grand canonical simulations with those of competing phases calculated by density functional theory, we predicted the phase diagram of the C-Ni-W ternary at multiple temperatures, which also, inherently showed the carbon solubility in Ni-W solid solution. A constrained ideal mixing approximation gives similar results, and can serve as a way to predict the phase diagram at lower temperatures. The presence carbon as impurity leads to a solid solution phase instead of the low temperature stable Ni$_8$W phase, perhaps explaining why that phase is not observed in experiment.

\bibliography{CNiW-prb}

\end{document}